# Optically Transparent Meta-Grating Embedded in Rear Windshields for Automotive Radar Detection

Sergey Geyman[1], Dmytro Vovchuk[1,*], Denis Kolchanov[1], Mykola Khobzei[1,2], Vladyslav Tkach[1,2], Hagit Gilon[1], Eyal Cohen[1], Eran Yunger[1], Vjačeslavs Bobrovs[1], Aviel Glam[1], Ofer Amrani[1], and Pavel Ginzburg[1]

[1]*Institute of Photonics, Electronics and Telecommunications, Riga Technical University, Riga, LV-1048, Latvia*
[2]*Department of Radio Engineering and Information Security, Yuriy Fedkovych Chernivtsi National University, Chernivtsi, 58002, Ukraine*

**Abstract**
Radar plays a crucial role in automotive safety by enabling reliable object detection, thereby assisting drivers and, prospectively, serving as one of the primary sensors in autonomous driving. The radar visibility of a road participant depends on its radar cross-section (RCS). While RCS is an inherent property, enhancing it, similar to using reflective vests for optical visibility, can significantly improve radar detection through cooperative target design. However, modern vehicles are not designed for this purpose, and embedded reflectors are not utilized due to the industry's conservative approach and the limited space available on the vehicle's exterior. Rear windshields offer a vast unused area, but they must still serve their primary function and remain transparent. We propose utilizing this area by embedding a reflecting surface that accounts for the interrogation scenario geometry and the angular tilt of the rear windshield, ensuring the wave is retroreflected back to the radar. The surface is realized as an array of thin conductive wires with a periodicity that provides in-phase excitation for the design incidence angle. Given that automotive radars operate in the millimeter-wave regime (77–81 GHz), large-scale surfaces with sub-millimeter manufacturing accuracy are required. This is achieved by imprinting conductive inks, composed of silver nanoparticles and binders, into grooves in the glass. The fabricated $10 \times 10$ cm$^2$ sample, with around 90% optical transparency, demonstrates an RCS of 8 m², surpassing the typical RCS of a car. Extrapolating this performance to the entire rear window with an embedded meta grating, a typical RCS of 1,000 m² can be achieved, thereby enhancing the detectability range by nearly an order of magnitude. Smart windows, which integrate multiple functionalities, including electromagnetic properties that do not compromise optical performance, enable advanced applications in wireless communication, such as automotive scenarios, IoT, and many others.

*corresponding author: Dmytro.Vovchuk@rtu.lv

**Introduction**

Ensuring road safety is a critical priority in modern transportation, with advanced driver assistance systems (ADAS) requiring the use of multiple sensors [1–5]. This need becomes even more crucial with the advancement of autonomous driving technologies, where reliable information is essential for governing machine-made decisions. Among the various sensing modalities, radar has emerged as a primary sensor due to its robustness in diverse environmental conditions, including low visibility, rain, and fog. Automotive radars operate primarily in the millimeter-wave range (77–81 GHz) and are used for collision avoidance, adaptive cruise control, and object detection [6–10].

Despite their advantages, automotive radars face significant challenges in achieving reliable detection and tracking of vehicles. The main one is the relatively poor angular resolution, which rarely falls below 1 degree. In addition, road traffic clutter, ground reflections, and severe multipath effects make target detection and classification challenging. A key factor influencing radar performance is the RCS of the target. This parameter determines an object's visibility to the radar. Unlike items designed to enhance optical visibility, such as reflective vests, vehicles are not inherently optimized for radar detection. For example, the RCS of passenger vehicles can range from approximately 10 to 20 dBsm (10 to 100 m²), depending on factors such as the vehicle's design, aspect angle, and surface materials [11–15]. For example, the ADAS radar from Texas Instruments (AWR2544) claims to detect passenger vehicles at distances of up to 200 meters [16]. The detectability and reliability of radar detection can be improved by incorporating cooperative targets designed with RCS enhancement, e.g., [17–20]. This approach is well established in maritime applications, where corner reflectors are mounted on top of yacht masts to ensure strong radar returns.

The profit from RCS enhancement can be estimated using the radar equation [21], which predicts that the received power follows $P_r \propto \sigma/R^4$, meaning it is directly proportional to the RCS ($\sigma$) and inversely proportional to the fourth power of the distance ($R^4$). Thus, improving a vehicle's reflectivity by boosting its RCS from 10 to 30 dBsm will increase the detection range by more than threefold, making the vehicle visible from well over 500 meters. Alternatively, it enhances the SNR at the receiver, enabling more reliable detection. It is also worth noting that cooperative targets [22–24] can encode reflections, granting radars advanced classification capabilities; however, this aspect is beyond the scope of this work.

Despite the clear advantages of radar cooperativity, mounting corner reflectors on vehicles remains a controversial approach, partly due to design-driven rather than engineering-driven decisions in the industry, where aesthetics are often prioritized over performance. In the endeavor to find space that is not occupied by other functional equipment, the rear windshield can be considered a viable option. In many cases, most of its area remains transparent and nonfunctional (the rear defogger, which consists of resistive wires for heating, is excluded from this discussion for the sake of clarity). Consequently, the proposed solution is to turn the rear windshield into a functional backscatter reflector, dramatically enhancing the RCS. The typical area of the rear windshield is about 1 m². As a rough initial estimate, one can assume normal incidence and apply the RCS formula for a square plate, given by: $\sigma = 4\pi A^2/\lambda^2$, where $A$ is the area of the square plate and $\lambda$ is the radar central wavelength. Substituting the numbers for 80 GHz carrier frequency, $\sigma = 60\ dBsm$ is obtained. While this number is enormously high and unlikely to be achieved in practice with a tilted window, it provides an upper bound, highlighting the potential effectiveness of this approach. Figure 1 provides a graphical illustration of the concept, demonstrating how stronger reflections enhance the SNR in radar detection.

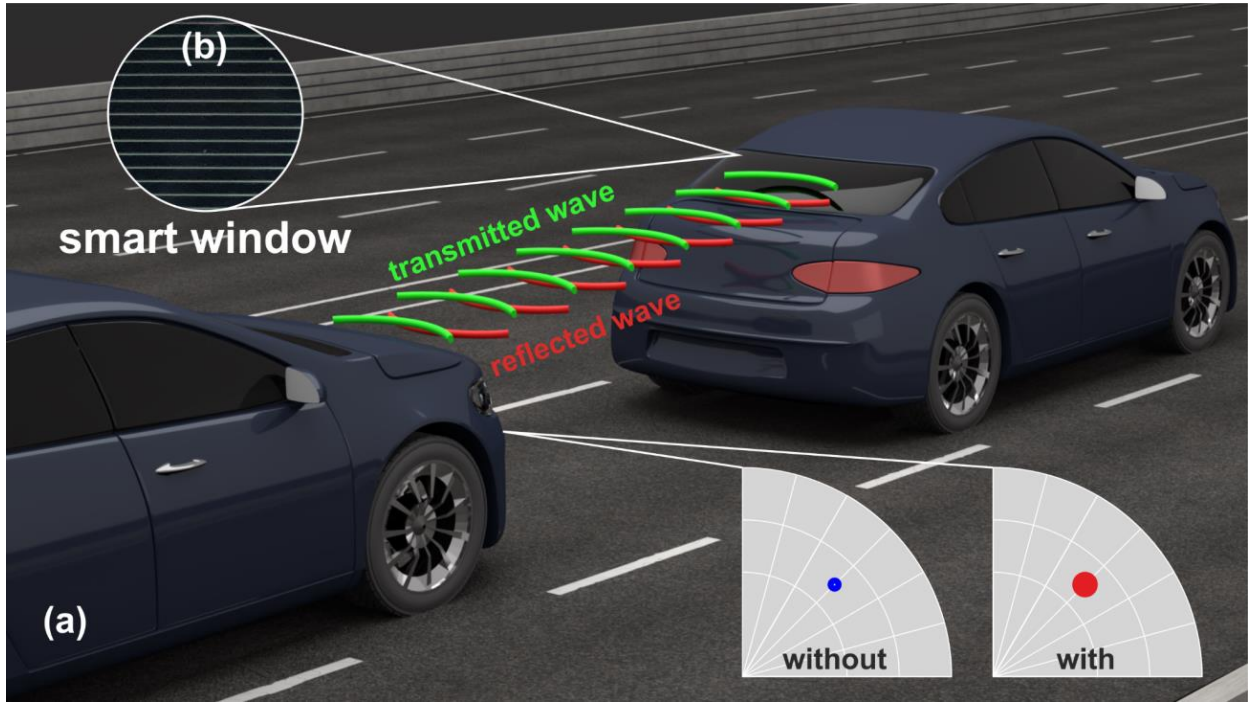

Figure 1. (a) Concept of a smart rear windshield demonstrating orders-of-magnitude enhancement in RCS, enabling efficient cooperation with ADAS. (b) Zoomed-in view of the system (meta grating on a glass, positioned on a black desk for visibility) showing the conductive wires that support strong backreflection.

Another important aspect to consider in this use case is the need to maintain optical transparency without compromising the primary function of the rear windshield. Optically transparent antennas and reflectors have gained significant attention due to their potential in various applications that require both electromagnetic functionality and visual transparency [25–28]. In the automotive industry, transparent antennas can be integrated into vehicle windshields to support wireless communication without obstructing the driver's view [29,30]. In wearable technology, particularly smart glasses and augmented reality devices, transparent antennas serve similar functions [31–33]. In homeland security applications, surfaces superimposed on camera apertures can block incoming bursts of radiation used as electromagnetic warfare countermeasures, preventing sensor saturation and maintaining operational integrity. This brief survey of use cases highlights the broader applicability of transparent antennas and reflectors beyond the specific use case we aim to explore.

Transparent antennas are typically realized using several key technologies, which are briefly reviewed here. Transparent Conductive Oxides (TCOs) are traditionally employed in electro-optical applications due to their ability to provide high optical transparency while maintaining good electrical conductivity at lower frequencies [34–36]. Indium Tin Oxide (ITO) is the most widely used TCO, known for its excellent electrical conductivity and high transparency. However, ITO has limitations, including brittleness and the high cost of indium. As a result, alternative materials like Aluminum-Doped Zinc Oxide (AZO) are being explored. AZO offers comparable transparency and conductivity to ITO but at a lower cost and with a greater abundance of raw materials. These properties make AZO a promising candidate for applications such as transparent electrodes in displays and photovoltaic devices [37]. Metal wire meshes serve as an alternative to TCOs, especially in applications requiring flexibility and durability. These meshes consist of fine metal lines patterned to create a conductive network with high optical transparency. For instance, NANOWEB® is a patented transparent conductive film made of an invisible, nanostructured metal mesh. It can be fabricated onto various surfaces, including glass and plastic, providing a combination of high transparency and conductivity [38]. Similarly, 3M™ Transparent Conductor Film utilizes a copper grid as the conductive layer, offering high light transmittance and low sheet resistance, making it suitable for applications like inconspicuous antennas [39]. We will use this technology to implement our concept in practice.

The report is organized as follows: it begins with the fabrication methodology, followed by a series of designs and experimental demonstrations. Specifically, polarization sensitivity is analyzed first, followed by the design and demonstration of a radar reflector, tailored to comply with rear windshield integration requirements. Finally, the entire concept is validated through a series of indoor and outdoor experiments using a commercial automotive radar.

**Fabrication of Large-scale Wire Line Array**

Among existing fabrication methods [40–42], Pattern Transfer Printing (PTP™) is to be mentioned as the one to be used to fabricate the samples in this report. PTP™ is based on laser-induced deposition from a polymer film that acts as a silver paste carrier. A desired trench geometry is embossed into a polymer film with a patterned rigid metal stamp, and the film is subsequently collected on rolls, which are then mounted in the PTP machine. The polymer-embossed roll acts as the carrier of the silver paste to be printed onto a receiving substrate. PTP™ enables printing of 10-30 μm line width [43].

Commercial PV paste (PV43A Solamet, Dupont) was mixed to achieve homogeneity by planetary mixer (Thinky SR-500, USA) for 4 min at 1200 rpm. Then, the paste was inserted into a print head and filled into the trenches using metal blades. The carrier segment is brought close to a 10 × 10 cm$^2$ glass. A high-power CW 1070nm laser beam illuminated the filled trenches through the transparent polymer film, resulting in a transfer of the paste from the embossed polymer film onto the glass, forming metallic lines. The main steps of the method [43] are presented in Figure 2(a). Figure 2(b) shows an optical image of the zoomed-out Wire Line Array, with straight lines printed uniformly across the area. The slight line curvature, measured on the millimeter scale, is still smaller than the radar wavelength, confirming that the quality meets the requirements of our application. This supports using this method instead of photolithography, as it offers better scalability and lower cost while providing the necessary quality without unnecessary complexity. Figure 2(c) shows a scanning electron microscope (SEM) image of a cross-sectioned printed line, revealing its internal structure. The silver nanoparticles and the granular texture are clearly visible. Despite this granularity, the lines remain electrically conductive, as verified by DC resistance measurements over 10 cm samples. The vendor-specific conductivity is $(1.7\text{-}2.0) \cdot 10^7$ S/m, which is also verified by our independent assessment. Later experimental demonstrations will also show that the lines perform well at high frequencies, meeting the millimeter wave requirements.

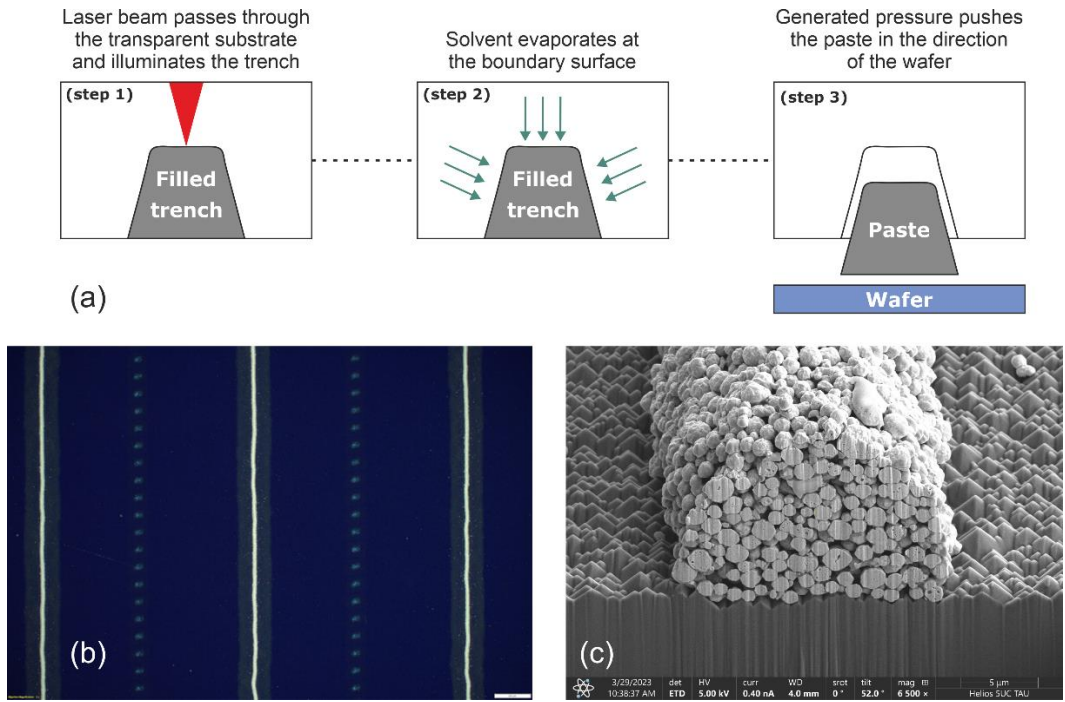

*Figure 2. (a) Schematic of the PTP™ fabrication steps. (b) Optical microscope image of a PTP-printed silver Wire Line Array on a photovoltaic silicon wafer, shown at 5× magnification. The white scale bar indicates 1 mm. (c) SEM image of the cross-section of a printed line.*

**Wire Line Array Polarizer**

The first and most straightforward test of a Wire Line Array (also called a meta grating) is its ability to exhibit a strong polarization-dependent response. An electromagnetic wave linearly polarized along the direction of the conductive wires should experience strong reflection due to efficient coupling with the structure. In contrast, a wave polarized perpendicular to the wires should interact only weakly with the array, rendering the sample nearly transparent to that polarization. For the proper operation of such a device, the periodic spacing between the wires must be smaller than half the wavelength within the operational band to avoid structural resonances and grating lobes. Accordingly, in this experiment, the distance between the wires was set to 0.15 mm.

The experimental setup appears in Figure 3(a). Two horn antennas are connected to the ports of a tandem-connected Keysight P9374A Network Analyzer, which drives an extender system enabling the measurement of complex-valued transmission and reflection coefficients across the 75–110 GHz frequency range, thereby fully covering the automotive radar band. The sample under test (SUT) is positioned 50 mm in front of the transmit antenna and is rotated with respect to the antenna polarization. The transmission coefficient ($|S_{21}|^2$) spectra are measured at each orientation. Figure 3(b) shows measurements at 0° (E-field is perpendicular to the conductive lines), 45° (diagonal), and 90° (parallel). The results demonstrate: (i) over 8 dB attenuation between 0° (blue curve) and 90° (yellow curve); (ii) the 0° closely matches the free-space (FS) response, indicating near-perfect transparency; (iii) the transmission characteristics exhibit minimal chromatic dispersion, as the wire spacing does not support lattice resonances across the entire frequency band. The difference between the blue and red curves is approximately 3 dB, which is consistent with the expected attenuation given by cos(45°). To achieve stronger isolation between polarizations, the spacing between the lines should be reduced. Since the wire thickness exceeds the electromagnetic skin depth and the cross section remains subwavelength, the response is effectively independent of radius. Our numerical verification indicates variations below 2 percent across a tenfold range of radii.

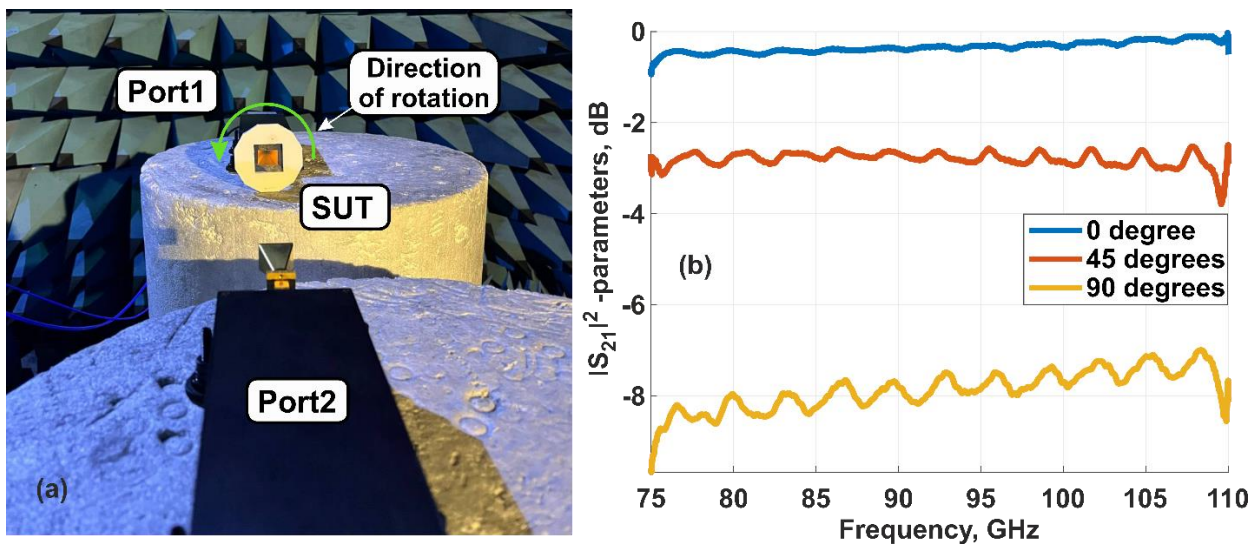

*Figure 3. (a) Photograph of the experimental setup inside the anechoic chamber. (b) Transmission power spectra of the polarizing sample for several orientations of the wire line array relative to*

*the wave polarization, as indicated in the legend, normalized to the transmission power spectrum of a free space.*

**Automotive Reflector**

In contrast to polarizer operation, where the spacing between wires must be small compared to the wavelength, the in-phase (up to a multiple of 2π) excitation of the wires is governed by a resonant mechanism described by the Floquet theorem [44]. In an array of infinite, parallel, thin, perfect electric conductor (PEC) wires, symmetry permits only Floquet modes whose propagation angles satisfy the grating equation: $sin\ \theta_m = sin\ \theta_i + m\lambda/d$, where $\theta_m$ is the angle of the m-th mode, $\theta_i$ is the angle of incidence, λ is the wavelength, m ∈ Z is the diffraction order, and d is the spacing between adjacent wires. To achieve retro-reflection for a given angle $\theta_i$, the period was chosen as:

$$2d sin\ \theta_i = \lambda, \tag{1}$$

so that only the modes with indices *m = 0* and *m = -1* remain propagating, while all higher-order harmonics become evanescent and therefore carry no power. Our design is tailored for automotive scenarios in which the rear windshield is tilted at an angle $\theta_i$ relative to the direction of incidence. Given long-range interrogation at distances of several hundred meters, variations in the angle between the radar position on the following vehicle and the rear windshield can be considered negligible.

Figure 4(a) shows the geometry of the Wire Line Array. The line period was tuned to satisfy the in-phase excitation condition calculated using Eq. 1. CST Microwave Studio was used for the analysis, employing the frequency-domain solver. Although our grid has finite dimensions, the principal predictions of infinite-array theory remain valid. The metal lines were modeled as PEC, a thin wire approximation. A 1 mm-thick glass layer was included as the substrate to match the subsequent experimental implementation. Periodic boundary conditions were not applied to verify that the results apply to a finite-sized structure. The inset in Figure 4 (a) reveals weak sidelobes adjacent to the dominant lobe, illustrating the presence of edge effects associated with the finite aperture. The far-field scattering diagrams are calculated for incidence angle $\theta_i$ = 45 degrees, frequency *f = 80 GHz*, structure period *d = 2.65 mm*, and array size 10 × 10 cm² (approximately 26.7 λ per side, i.e., 38 periods). In this case, a strong retroreflection and three other lobes are observed, in full agreement with the Floquet theorem. Each propagating order m generates two modes: one transmitted through the grid and one reflected. The *m = 0* order produces the familiar specular pair, with reflected and transmitted beams at the same 45° angle to the normal. The *m = −1* order gives the remaining two waves: the reflected beam is directed exactly back toward the source (retroreflection), while its transmitted counterpart emerges symmetrically into the forward half-space.

To assess the angular and frequency sensitivity of the 10 × 10 cm² sample with a 2.65 mm period, its RCS spectra were modeled. The results, presented in Figure 4(b), show the RCS colormap, where each vertical slice represents a spectrum with a strong maximum corresponding to the in-phase excitation condition, as can be verified by applying Eq.1. As shown in the results, the 10 × 10 cm² sample already demonstrates an RCS of 8 m², which is comparable to that of a private vehicle. Considering the scaling relation $\sigma = 4\pi A^2/\lambda^2$, a dramatic enhancement can be achieved if the Wire Line Array covers the entire rear window. Additionally, the estimated half-width of the scattered lobe is 2.7°, which makes the system tolerant to positioning inaccuracies between the radar and the sample. This condition can be relaxed by introducing aperiodicity, although it comes at the expense of RCS performance. Considering that fabrication uses pattern transfer printing, assessing tolerance to a possible conductivity reduction is valuable. The inset in Fig. 4(b) shows the backward radar cross section as a function of frequency for several conductivity values at an incidence angle of 45°. Across a span of seven orders of magnitude in conductivity, the radar cross section changes by at most one order of magnitude, indicating robust performance. The explored conductivity values are well below the vendor specification; therefore, the high conductivity approximation used in the modeling is justified.

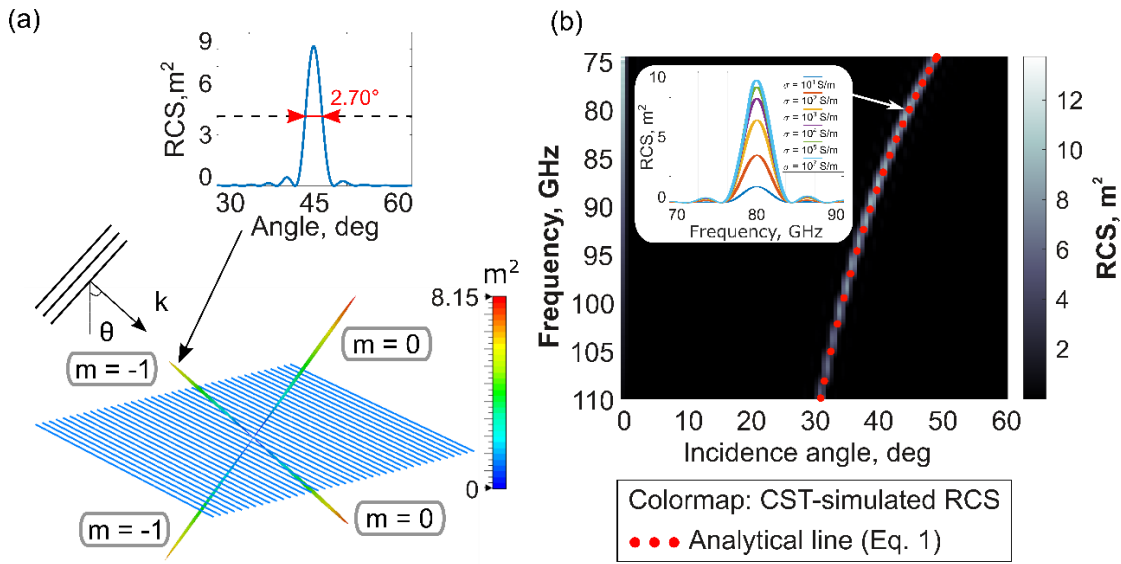

Figure 4. *Wire Line Array performance based on numerical results obtained in CST Microwave Studio. (a) Scattering pattern of the array illuminated by a plane wave at 80 GHz with an incidence angle of 45°. Inset: profile of the scattering lobe. (b) RCS colormap in logarithmic scale as a function of frequency and incidence angle for a plane wave. The red dots indicate the* in-phase excitation *condition for first-order backscattering (Eq. 1). Inset: Backward radar cross section versus frequency for multiple conductivity values.*

The experimental samples were constructed according to the design specifications and using the technique described in the 'Fabrication' section. The reference sample was assembled as an array of stretched copper wires, which were affixed to adhesive paper. The reference sample serves to validate the electromagnetic performance of the additively manufactured sample on glass. Figure 5(a-c) demonstrates the setup and the experimental samples. In this set of experiments, the reflection coefficients were measured. For this purpose, the transmitting and receiving antennas were co-located, in contrast to the previous transmission setup used for validating the polarizer performance. Angle-dependent reflection measurements were performed in an anechoic chamber, as shown in Figure 5(a), to validate the strong reflection at the in-phase excitation condition and confirm agreement with the numerical analysis. The experimental setup is fully controlled via MATLAB on a PC, with the sample under test (SUT) placed on a white styrofoam cylinder at the center of a rotation table; the styrofoam is transparent to GHz waves. Two types of SUTs were measured: metal wire arrays serving as a reference sample with a size of $10 \times 10$ cm$^2$, wire diameter of 0.02 mm, and a lattice period of 2.65 mm (Figure 5(b)); and samples with conductive wires printed on a glass substrate (Figure 5(c)). The measurements were performed with an azimuthal rotation precision of 1º over a range from 0º to 60º, with the conductive wires aligned parallel to the vertically polarized antenna.

As a result, strong reflection is observed across the entire frequency range of 75 to 110 GHz at 0º, with only a slight minimum near 90 GHz. Additionally, the reflection peak appears at angles between 30º and 50º, depending on the wavelength, due to the in-phase excitation condition, designed for 45º. Figure 5(d) summarizes the experimental results for the reference sample, while panel (e) shows the performance for the Wire Line Array on glass. Although the reference sample aligns more closely with the basic theory and outperforms the glass-based structure, both results are in good agreement. The differences can be attributed to two main factors. First, the glass substrate introduces minor ohmic losses (in automotive glasses, loss tangent (*tan(δ)*) can vary between 0.01 and 0.02 at 80 GHz), while the reference sample does not use a substrate. Second, there are differences in electrical properties. The nickel wires in the reference sample closely approximate a perfect electric conductor, whereas the silver paste used on glass may introduce additional losses. However, this degradation is relatively minor and does not significantly affect the overall device performance, supporting its use as a radar reflector and confirming the feasibility of subsequent radar testing.

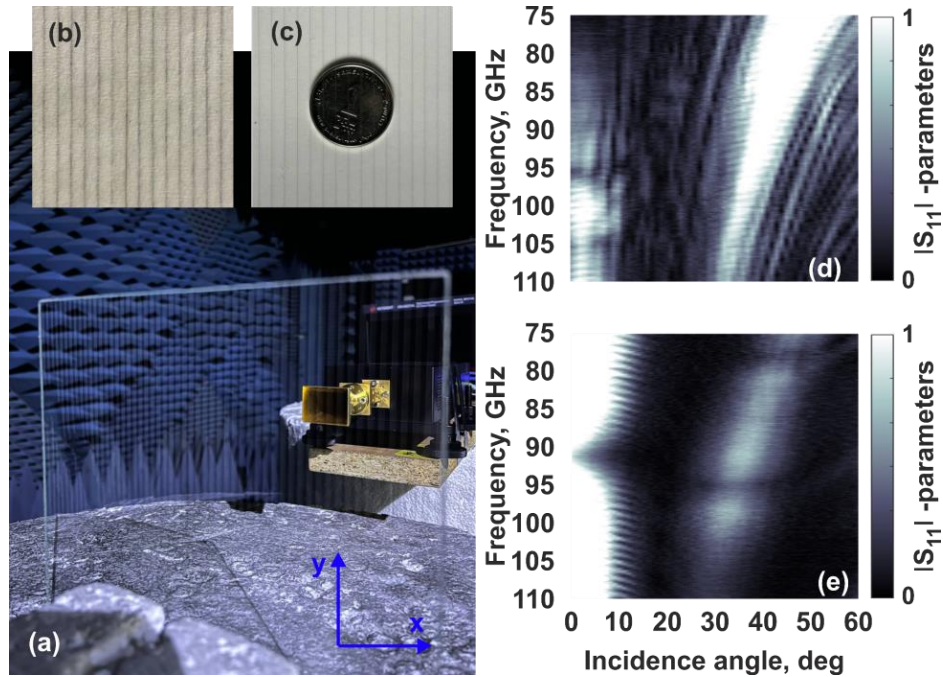

*Figure 5. Validation of the in-phase excitation condition in Wire Line Arrays. (a) Experimental setup inside an anechoic chamber. Samples: (b) reference sample with stretched copper wires, and (c) wire line array on windshield glass. For visualization, the glass sample is placed on paper to reveal the conductive wires; otherwise, it appears transparent as seen in panel (a). A small coin is placed on top to indicate the scale. The antenna polarization is vertical, and so are the metallic wires. (d, e) Reflection coefficient colormaps as a function of sample rotation angle (horizontal axis) and frequency (vertical axis). Strong reflection is observed at 0º (normal incidence) and between 30º and 50º, corresponding to the in-phase excitation condition. Panel (d) shows the reference sample, while panel (e) shows the glass windshield sample.*

**Optical properties**

The optical properties of the on-glass Wire Line Array were investigated using a standard optical transmission setup, as shown in Figures 6(a) and 6(b). A white light beam from an Avantes (AVALIGHT-HAL-S-MINI) lamp was collimated using a lens, ensuring that the sample was illuminated by a uniform circular spot with an area of approximately 180 mm². This setup ensures that more than eight lines fall within the illuminated area on the sample (for reference, the automotive reflector used in the initial electromagnetic study had a line spacing of 2.75 mm on a glass substrate and 0.15 mm on a flexible polyethylene terephthalate (PET) substrate). The transmitted light was then collected into an optical fiber using a focusing lens. The transmission spectrum was recorded with an Avantes (AvaSpec-ULS2048L) spectrometer and normalized to the lamp signal measured without the sample.

To evaluate potential transparency degradation, the optical transmission of the plain glass substrate was compared with that of the Wire Line Array-deposited glass. As shown in Figure 6(c), the bare glass exhibits an average transmission of approximately 90% in the visible range, which aligns with standard values for plain glass (ASGS reference). The deposition of the automotive reflector structure, comprising widely spaced conductive lines, results in a minimal transmission drop of around 2%, indicating negligible optical loss and excellent transparency retention.

To further demonstrate the versatility of the fabrication method, the Wire Line Array was also implemented on a PET film with a thickness of 100 µm. Figure 6(d) presents the transmission characteristics of this PET-based automotive reflector. In this case, the average transmission is reduced by approximately 4.6% across the visible range. This more pronounced decrease is attributed to the inherently lower transparency of the PET substrate, and possible thermal hazing effects introduced during deposition may further contribute to the loss. Importantly, in both glass and PET samples, the transmission reduction is spectrally flat, suggesting the absence of resonant optical phenomena and confirming that the observed losses are primarily due to geometrical and material effects rather than interference or plasmonic resonances.

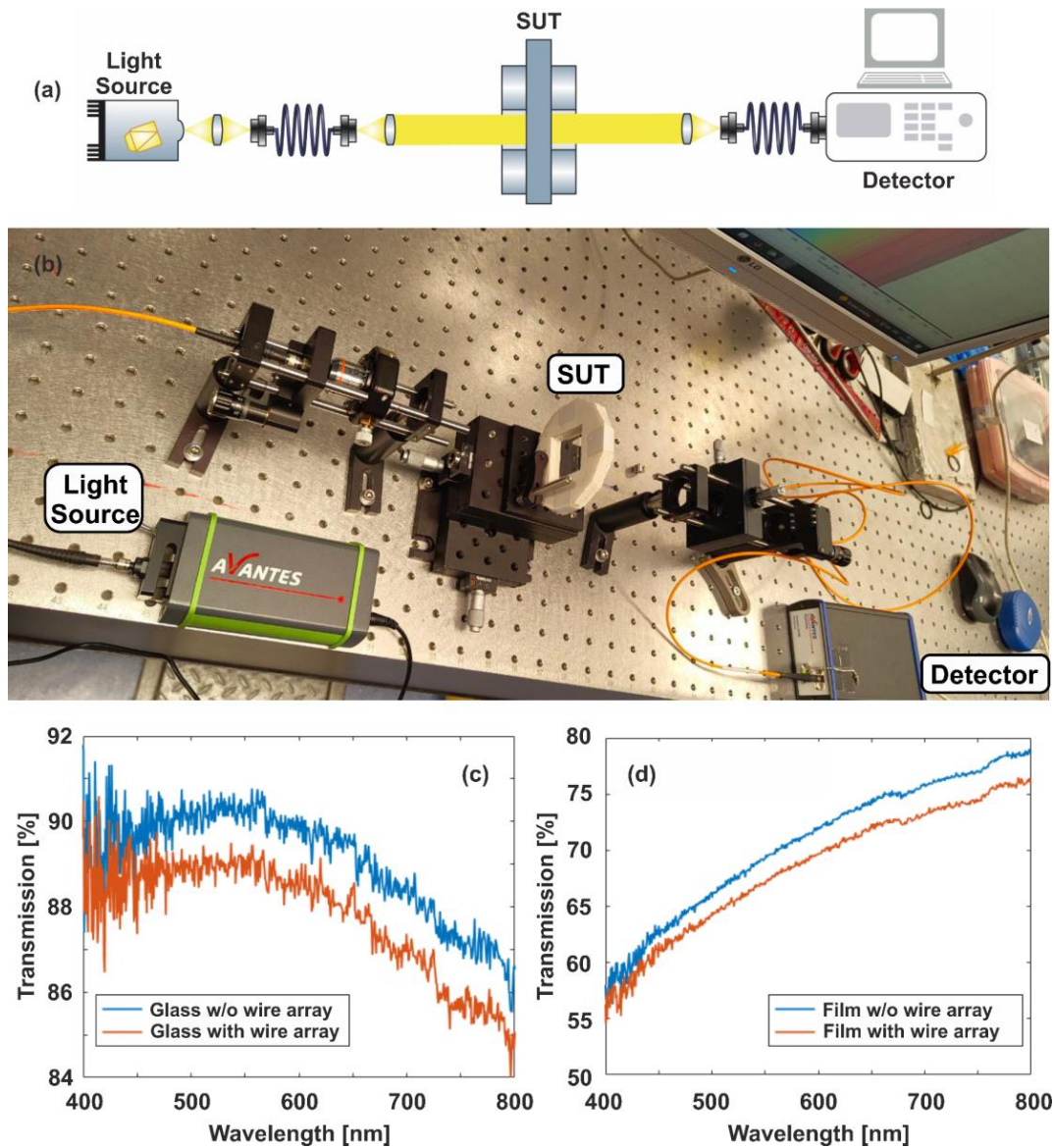

*Figure 6. Optical transparency assessment. (a) Schematic of the experimental setup. (b) Photograph of the measurement arrangement. (c, d) Transmission spectra of the wire line array on glass and a polymer film substrate, along with reference samples, as indicated in the legends.*

**Validation with TI AWR1642 automotive radar**

*Indoor measurements, static target*

The experimental setup consists of a TI AWR1642 automotive frequency modulated continuous wave (FMCW) radar operating at the 77-81 GHz band and two types of samples: a dihedral corner reflector (10 × 20 cm) and the Wire Line Array on glass (10 × 10 cm$^2$ in size) (Figure 7a). The electric field is vertically polarized, aligned parallel to the wires. The radar is controlled via

MATLAB, enabling the collection of object coordinates, relative backscattered power, and range information. In subsequent experiments, Doppler shift measurements will also be performed. The samples are positioned 1.5 meters away from the radar at the center of a rotation table, allowing measurements at different incidence angles ranging from 0° to 60°. In contrast to the RCS measurements, the radar output is the relative received power, measured per range and velocity bin. In the anechoic chamber, strong reflections from the object are clearly identifiable within the radar's target range bin.

Figure 7(b) shows the relative received power from the range bin as a function of the rotation angle of the target relative to the radar. The corner reflector, used as a reference, exhibits uniform backscattering up to an incidence angle of 40°, with an additional peak near 45°, which is characteristic of a dihedral corner reflector (Figure 7(b), blue solid line). Since the commercial corner reflector available for the experiment has an aperture twice the size of the Wire Line Array, a 6 dB correction was applied to account for the fourfold increase in RCS with doubling the area, as shown by the red dashed curve. The same measurement procedure was applied to the wire array, and the results are presented as the yellow curve.

Two conclusions can be drawn from the obtained results: (i) The scattering performance of the corner reflector prevails over the line wire array by 10 dB, which is rather expected, as the corner reflector provides the upper bound for RCS in case of electrically large nonresonant structures. (ii) The angular dependence of scattering from the wire line array follows the pattern observed in Figure 5(g) at the 80 GHz central frequency. Specifically, the scattering maximum occurs for incidence angles between 0° and 8°, followed by a 20 dB drop, with a scattering boost at 48° (a small shift from the CST model), corresponding to the in-phase excitation condition.

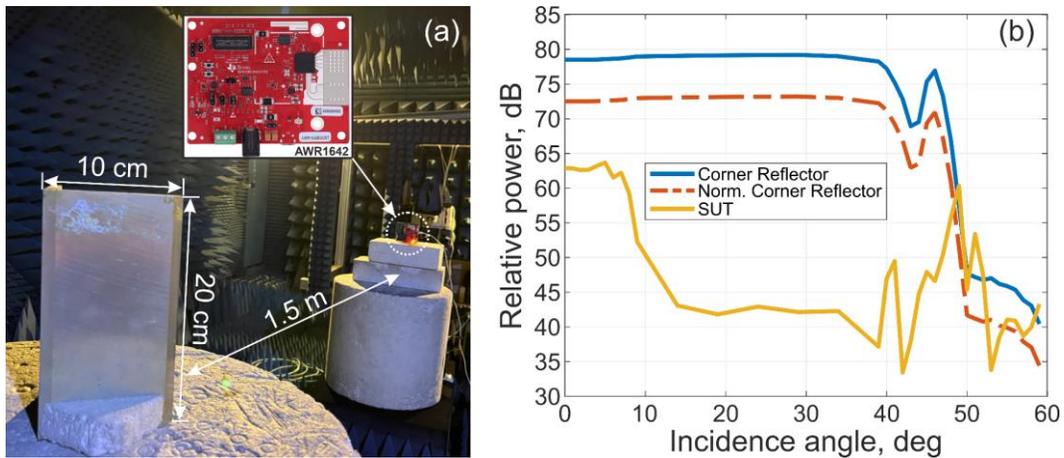

Figure 7. (a) Photograph of the indoor experimental setup using the AWR1642 automotive radar. The electric field is vertically polarized, aligned parallel to the wires. (b) Angular dependence of the relative power of the radar's received signal for both the corner reflector (blue) and the Wire Line Array (yellow). To match the sample areas, the corner reflector aperture was scaled to the size of the array, and the interpolated response is shown as a red dashed line.

*Indoor measurement, target in motion*

To assess the impact of motion, the experimental setup utilizes the AWR1642 automotive radar in conjunction with a conveyor belt that enables controlled target translation (Figure 8). The radar interface, controlled via MATLAB, allows for the tuning of parameters such as chirp and frame configurations, and enables the application of clutter removal filters. The visualization output can be adjusted to display range, Doppler, coordinates, and relative peak power. Simultaneously, MATLAB controls the conveyor's velocity, initial and final positions, and start timing. The conveyor is positioned in front of the radar at a distance of approximately 1.4 to 1.5 meters.

In this set of experiments, two targets were assessed: a corner reflector (Figure 8(a)) and the Wire Line Array (SUT), tilted at 48° (Figure 8(d)).

Figures 8(b) and 8(e) show the target trajectories. The polar plots display the radar output, where targets at different range bins and azimuth angles appear as dots. The elevation is fixed at 0°, and the entire interrogation scenario occurs within a single plane. The targets moved along straight lines, with their positions sampled over the entire trajectory, which is visible in the plots. From this set of experiments, it can be observed that the higher RCS target (corner reflector) is accurately detected along the entire range of conveyor motion at a 90° azimuth angle. In contrast, the Wire Line Array exhibits some detection ambiguity at larger distances. This directly demonstrates how higher RCS improves the SNR in detection, thereby enhancing radar performance.

The insets to panels (b) and (e) show the distribution of reflected relative power peaks as a function of distance. Those are the screenshots from the TI online demo version that confirm the utilized algorithm. The observed several-centimeter positioning accuracy, though not directly representing resolution, aligns well with the 4 GHz bandwidth of the FMCW radar.

To examine the impact of clutter, radar output intensities were recorded using its standard, uncustomized interface. For each of the ten different radar range bins along the conveyor belt, the intensity was recorded over a time window of several seconds, with each point in Figure 8(c) corresponding to one of these measurements. It can be seen that despite the controlled environment of the chamber, the intensity fluctuates within a range of approximately 10 dB. Additionally, the absolute values of the received signals show that the corner reflector produces a signal approximately 15 dB stronger than that of the Wire Line Array, in full agreement with the results presented in Figure 7(b). To address the clutter issue, a Doppler filter was applied to track only the moving targets. Figures 8(f) and 8(g) show the extracted velocities for both targets, demonstrating a very high level of accuracy that closely matches the ground truth obtained from the conveyor belt controller. In both cases, regardless of the differences in RCS, the measured velocities remained stable without fluctuations. The results were obtained by analyzing the motion over a distance from 1.45 m to 1.85 m, with a ground truth velocity of 6 mm/s. Detecting such a slow movement is not trivial, making this measurement accuracy particularly notable.

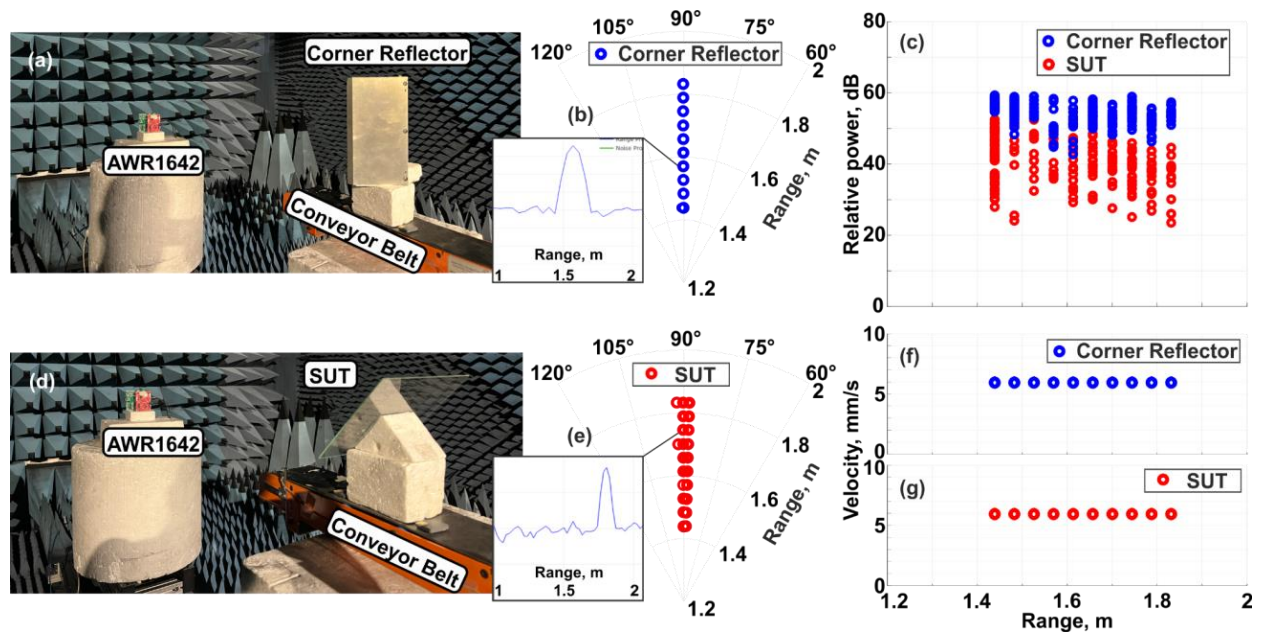

*Figure 8. Target assessment in motion. (a) Corner reflector and (d) Wire Line Array (SUT) placed on a conveyor belt inside the anechoic chamber. The data was collected by moving the targets along a straight-line trajectory on the conveyor while interrogating them with the TI AWR1642 automotive radar. (b, e). Target trajectories during the motion. The polar plots display radar output, where targets at different range bins and azimuth angles appear as dots. The elevation is fixed at 0°, and the measurements occur within a single plane. The targets move along straight lines, with positions sampled over the entire trajectory. Insets show the reflected energy distribution along the interrogation line (for a selected position), with screenshots taken from the TI online demo interface. (c) Reflected signal intensity fluctuations at different conveyor positions. (f) and (g) Extracted velocities at different distances over the 1.45–1.85 m path, confirming the 6 mm/s ground truth.*

*Outdoor measurement*

The outdoor measurements are performed with the same radar, using the wire line array (1), corner reflector (2), and car (3) as samples (Figure 9(a)). The radar output from the vendor interface is used for data acquisition, with all radar settings matching those used during the indoor measurements. Figures 9(a) and (d) are photographs demonstrating the arrangement of the detected objects, showing the car, corner reflector, and wire line array on glass. The difference between the scenes is the presence of the car, which corresponds to the results presented in the first row of panels, and its absence in the lower row. The relative arrangement of the targets was set to ensure that the reflected power from the car, corner reflector, and the Wire Line Array remained at comparable levels, achieved by adjusting their distances to the radar in accordance with the fourth-power distance dependence in the radar equation. Figures 9(b) and (e) represent the scatter plot as generated by the radar software. The main targets are clearly visible, with a few other points corresponding to the clutter. The relative detection threshold can be further adjusted but was set to match the requirements of the current scenario. Figures 9(c) and (f) show the relative received power as a function of distance (range gate). A single radar beam, directed towards the targets, was considered (the beam can be selected in the software). In the first scene, all three targets are clearly visible. The corner reflector appears stronger, as it is positioned twice as close, giving it an additional 12 dB advantage over the car. In scene 2, the wire line array shows a 20 dB smaller reflection, with an additional penalty due to the distance change, as observed in the previous case. Overall, these results correspond well with accurate measurements in Figure 7(b). Indoor and outdoor experiments confirm that the rear windshield enhancement concept using the Wire Line Array is effective, demonstrating its potential for practical automotive radar applications.

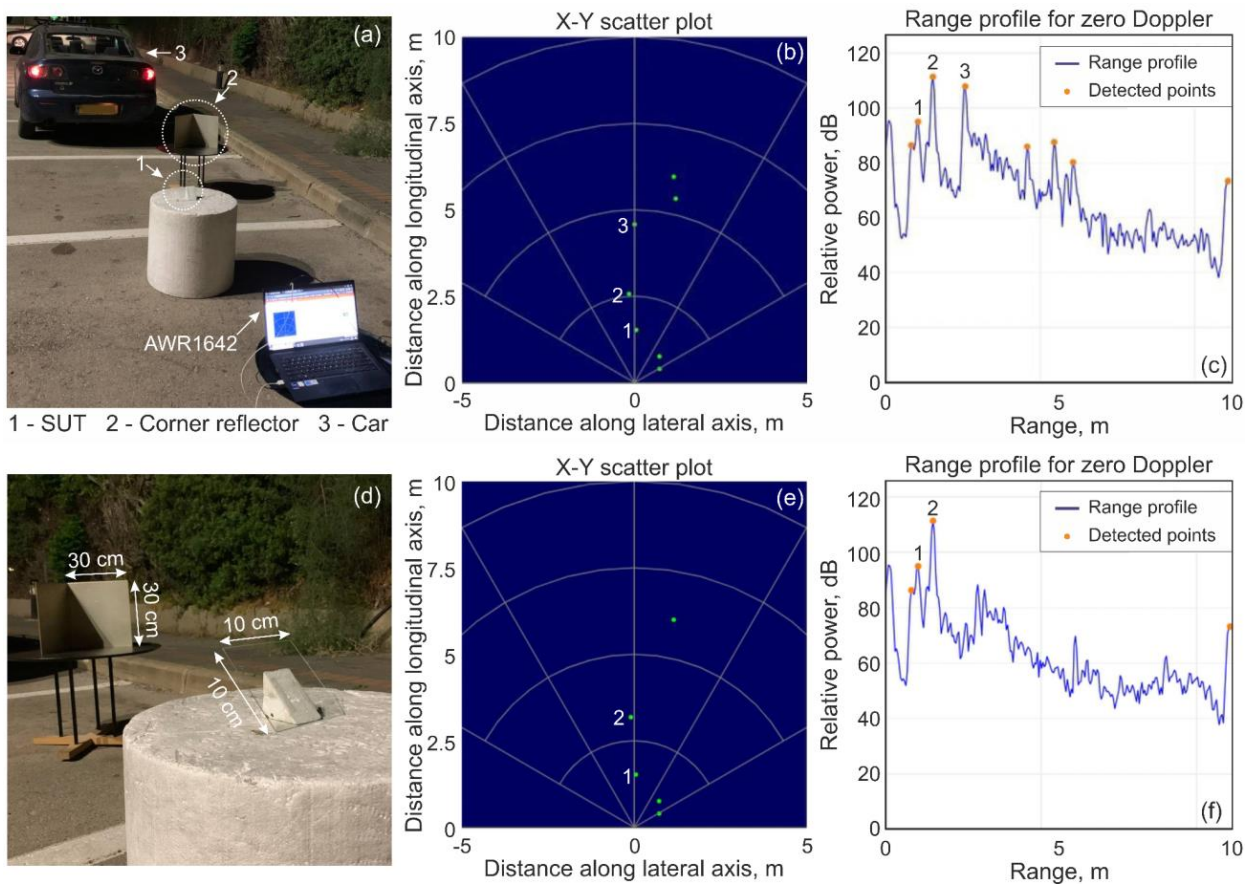

Figure 9. Outdoor Experiments (a), (d) Photographs of the outdoor scenes, where (1) is the wire line array, (2) is the corner reflector, and (3) is the car. (b), (e) Corresponding scatter plots as displayed in the radar software. (c), (f) Relative reflected power as a function of the range to the target. The central radar beam was selected.

## Conclusion

The challenge of enhancing the millimeter-wave RCS of vehicles to improve detection by automotive radars, thereby increasing driving safety, was addressed. This was achieved by integrating transparent wire line arrays onto rear windshields without adding bulky components or altering vehicle geometry, thus maintaining conformity and preserving optical transparency and mechanical integrity. To address this need, a scalable and cost-effective approach was developed using wire line arrays fabricated by PTP on both glass and flexible polymer substrates. Arrays with a period of 2.65 mm and line widths of 20 μm were fabricated over areas of up to $10 \times 10$ cm$^2$. Optical measurements confirmed high transparency with less than a 2% reduction in visible transmission, while electromagnetic simulations and experiments demonstrated effective polarization control and strong reflections across the 75–110 GHz frequency band. The practical effectiveness of the structures was validated using radar experiments with the TI AWR1642 module, where clear target detection, enhanced reflected signals, and stable Doppler-based velocity tracking even at low speeds were observed. The concept was validated in both indoor and outdoor scenarios, confirming that RCS enhancement can be achieved while preserving transparency. Extrapolating the demonstrated concept to a full rear windshield could achieve RCS values up to 1000 m², potentially extending reliable detection ranges beyond 500 m, significantly enhancing SNR and detection reliability in automotive radar systems.

The demonstrated technology can be extended to other transparent vehicle surfaces, including windshields, side windows, and sunroofs, to enhance radar cooperation for ADAS and V2X (Vehicle-to-Everything) systems. Beyond automotive applications, similar metasurface-based approaches could be utilized in infrastructure-to-vehicle (I2V) communication, transparent signage for smart transportation systems, and drone landing platforms requiring stealth-to-visible transitions under millimeter-wave radar interrogation. Furthermore, the concept can be developed into a functional antenna integrated within the transparent structure, enabling dual-use for sensing and communication. The scalability and low-cost fabrication demonstrated here open a pathway for integrating advanced electromagnetic functionalities into transparent surfaces across multiple use cases.

## Data Availability Statement

The article's data will be shared on a reasonable request from the corresponding author.

## Acknowledgement

The Riga Technical University and Yuriy Fedkovych Chernivtsi National University's teams acknowledge the support provided by the Horizon Europe project "Towards an Excellence Centre on Quantum Photonics in Latvia" (Grant Agreement ID: 101160101).

**Conflict of Interest Disclosure**

The authors declare no conflict of interest.